\documentclass[aps,prd,showpacs,showkeys,amsmath,amssymb,11pt]{revtex4}
\usepackage{amssymb}
\usepackage{amsmath}
\usepackage[dvips]{graphicx}
\usepackage{latexsym}

\newcommand{\be}{\begin{equation}}
\newcommand{\ee}{\end{equation}}
\newcommand{\ba}{\begin{array}}
\newcommand{\bqa}{\begin{eqnarray}}
\newcommand{\eqa}{\end{eqnarray}}
\newcommand{\cO}{{\cal O}}
\newcommand{\mL}{\mathcal{L}}

\begin{document}

\begin{flushright}
{IFIC/08-32}
\end{flushright}

\title{Study of $\tau^- \to V P^- \nu_\tau$ in the
framework of resonance chiral  theory}

\author{ Zhi-Hui~Guo}

\affiliation{\small \mbox{ Department of Physics,
Peking University, Beijing 100871, P.~R.~China, }  \\
 \mbox{ and IFIC, CSIC-Universitat de Val\`encia,
 Apt.~Correus 22085, E-46071, Val\`encia, Spain} }

\begin{abstract}
In this paper we study two kinds of $\tau$ decays: (a) $\tau^-\to(
\rho^0\pi^-,\omega\pi^-,\phi\pi^-,K^{*0}K^-)\nu_\tau$, which belong
to $\Delta S=0$ processes and (b) $\Delta S = 1$ processes, like
$\tau^- \to (\rho^0 K^-,\omega K^-,\phi K^-,\overline{K}^{*0}\pi^-
)\nu_\tau$, in the framework of resonance chiral theory (R$\chi$T).
We fit the $\tau^- \to \omega \pi^- \nu_\tau$ spectral function and
the invariant mass distribution of $\omega K$ in the process of
$\tau^- \to \omega K^- \nu_\tau$ to get the values of unknown
resonance couplings. Then we make a prediction for branching ratios
of all channels.
\end{abstract}

\maketitle

\section{Introduction}

As the only lepton with the ability of decaying into hadrons, $\tau$
decay provides an excellent environment to study the nonperturbative
dynamics of QCD. In these decays, the intermediate resonances may
play an important role. On the other side, due to the improvement of
statistically significant measurements, the branching ratios and
spectral functions of the processes containing resonances in the
final states of $\tau$ decays have also been determined in recent
experiments
\cite{pdg}\cite{expwpi}\cite{expwk}\cite{expphikbelle}\cite{expphikbabar}.
Some theoretical discussions have been devoted to the study of a
$\tau$ decaying into a resonance plus a pseudo-Goldstone meson and a
tau neutrino in literature. Angular decay distribution for
$\tau\rightarrow \omega\pi\nu_\tau$ was studied in
\cite{angularwpi}.
 A chiral lagrangian derived in a similar way to the Nambu Jona-Lasinio model,
has been used to investigate mesonic $\tau$ decays in
\cite{theotauvpli}. Vector meson dominance model has also been
applied to the studies of  $\tau$ decaying into $\phi$($\omega$)
plus  one pseudoscalar meson \cite{theotauvpmex} recently.

 Chiral perturbation theory ($\chi$PT) is a powerful tool to
describe the interaction for pseudo-Goldstone mesons, which is based
on the chiral symmetry and the momentum expansion \cite{gl845}. When
the typical energy scale of the process reaches around $M_\rho$,
$\chi$PT has been extended to resonance chiral theory (R$\chi$T),
where all of the symmetry allowed operators, consisting of specific
number of multiplets of resonances and pseudo-Goldstone mesons, can
be introduced in a systematic way \cite{rchpt89}\cite{2reson}\cite{3reson}.
To build a more realistic QCD-like effective theory, large-$N_C$ techniques
and short-distance constraints from QCD have  been implemented into
the resonance effective theory to constraint resonance couplings
\cite{moussallam}\cite{knecht}\cite{vvp}\cite{vap}. Therefore,
resonance chiral effective theory can be a perfect tool to study
hadronic $\tau$ decays. Indeed it has already been employed in the
studies of $\tau \rightarrow \pi K \nu_\tau$ \cite{taupik}, $\tau
\rightarrow \pi\pi\pi\nu_\tau$ \cite{tau3pi} and $\tau \rightarrow
K\overline{K}\pi\nu_\tau$ \cite{tau2kpi}.

In this paper, we will make a comprehensive analysis for $\tau$
decaying into a vector resonance plus a pseudo-Goldstone meson and a
tau neutrino: (a) $\Delta S =0$ processes, such as $\tau^-
\rightarrow ( \rho^0\pi^-, \omega\pi^-, \phi\pi^-, K^{*0}K^- )
\nu_\tau$ and (b) $\Delta S = 1$ processes, like $\tau^- \rightarrow
( \rho^0 K^-, \omega K^-, \phi K^-, \overline{K^{*0}}\pi^-)
\nu_\tau$, in the frame of R$\chi$T.

\section{ Theoretical frame for tau decays }

The amplitude for $\tau^-(p) \rightarrow P^-(p_1) V(p_2)
\nu_\tau(q)$, where $P^-$ can be $\pi^-, K^-$ and $V$ can be
$\rho^0, \omega, \phi, K^{*0}, \bar{K^{*0}}$, has the general
structure \bqa\label{general} -G_F V_{uQ} \overline{u}_{\nu_\tau}(q)
\gamma^\mu(1-\gamma_5)u_{\tau}(p) [\,\, v
\,\varepsilon_{\mu\nu\rho\sigma} p_1^{\rho} p_2^{\sigma} - ( a_1 \,
g_{\mu\nu} + a_2 \, {p_1}_\mu {p_1}_\nu +  a_3 \, {p_2}_\mu
{p_1}_\nu )
  \,\,] \epsilon^{*\nu}_V(p_2) \,,
\eqa where $G_F$ is the Fermi constant; $V_{uQ}$ is the CKM matrix
element; $\varepsilon_{\mu\nu\rho\sigma}$ is the anti-symmetric
Levi-Civit\`a tensor; $\epsilon^{*\mu}_V(p_2) $ is the polarization
vector for the vector resonance; $v$ denotes the form factor of the
vector current
 and $a_1, a_2, a_3$ are the corresponding axial-vector form factors.

We will evaluate the form factors in Eq.(\ref{general}) using R$\chi$T.
 The leading $O(p^2)$ lagrangian of $\chi$PT is
\bqa \mL_2=\frac{F^2}{4}\langle u_\mu u^\mu + \chi_+ \rangle \,.
\eqa The kinematic term for the  (axial) vector resonance, in the
antisymmetric tensor formalism, is
\be
\mathcal{L}_{kin}(R)=-\frac{1}{2}<\nabla^\lambda
R_{\lambda\mu}\nabla_\nu
R^{\nu\mu}-\frac{1}{2}R^{\mu\nu}R_{\mu\nu}>,\,\,\,\, R=V,A
\ee
and the relevant interaction lagrangian only including one multiplet of
resonances are given by \cite{rchpt89} \bqa
\mL_{2V}=\frac{F_V}{2\sqrt2}\langle V_{\mu\nu} f_+^{\mu\nu}\rangle +
\frac{i G_V}{2\sqrt2}\langle V_{\mu\nu}[ u^\mu, u^\nu]\rangle \,,
\eqa \bqa \mL_{2A}=\frac{F_A}{2\sqrt2}\langle A_{\mu\nu}
f_-^{\mu\nu}\rangle \,. \eqa The interaction operators containing
two multiplets of resonances have also been written down in
\cite{vvp}\cite{vap}: \bqa \mL_{VVP}=&& d_1
\varepsilon_{\mu\nu\rho\sigma} \langle \{V^{\mu\nu},
V^{\rho\alpha}\} \nabla_{\alpha}u^{\sigma}  \rangle + i d_2
\varepsilon_{\mu\nu\rho\sigma} \langle \{V^{\mu\nu},
V^{\rho\sigma}\} \chi_- \rangle \nonumber \\ && + d_3
\varepsilon_{\mu\nu\rho\sigma} \langle \{ \nabla_\alpha V^{\mu\nu},
V^{\rho\alpha}\} u^{\sigma}  \rangle + d_4
\varepsilon_{\mu\nu\rho\sigma} \langle \{ \nabla^\sigma V^{\mu\nu},
V^{\rho\alpha}\} u_{\alpha}  \rangle \,, \eqa \bqa \mL_{VJP}=&&
\frac{c_1}{M_V} \varepsilon_{\mu\nu\rho\sigma} \langle \{
V^{\mu\nu}, f_+^{\rho\alpha} \} \nabla_{\alpha}u^{\sigma}  \rangle
+\frac{c_2}{M_V} \varepsilon_{\mu\nu\rho\sigma} \langle \{
V^{\mu\alpha}, f_+^{\rho\sigma} \} \nabla_{\alpha}u^{\nu}  \rangle
 \nonumber \\ && + \frac{i c_3}{M_V} \varepsilon_{\mu\nu\rho\sigma}
 \langle \{ V^{\mu\nu}, f_+^{\rho\sigma}\} \chi_- \rangle
+\frac{i c_4}{M_V} \varepsilon_{\mu\nu\rho\sigma} \langle V^{\mu\nu}
[ f_-^{\rho\sigma}, \chi_+ ] \rangle \nonumber \\ &&+
\frac{c_5}{M_V} \varepsilon_{\mu\nu\rho\sigma} \langle
\{ \nabla_{\alpha} V^{\mu\nu}, f_+^{\rho\alpha}\} u^{\sigma}  \rangle
+ \frac{c_6}{M_V} \varepsilon_{\mu\nu\rho\sigma} \langle
\{ \nabla_{\alpha} V^{\mu\alpha}, f_+^{\rho\sigma}\} u^{\nu}  \rangle
\nonumber \\ &&+ \frac{c_7}{M_V} \varepsilon_{\mu\nu\rho\sigma}
\langle \{ \nabla^{\sigma} V^{\mu\nu}, f_+^{\rho\alpha}\} u_{\alpha}  \rangle \,,
\eqa
\bqa\label{lagvap}
\mL_{VAP}=&& \lambda^{VA}_1 \langle [ V^{\mu\nu}, A_{\mu\nu} ] \chi_- \rangle +
i \lambda^{VA}_2 \langle [ V^{\mu\nu}, A_{\nu\alpha} ] h^\alpha_\mu \rangle +
i \lambda^{VA}_3 \langle [ \nabla_\mu V^{\mu\nu}, A_{\nu\alpha} ] u^\alpha
\rangle+ \nonumber \\&& + \,
i \lambda^{VA}_4 \langle [ \nabla^\alpha V^{\mu\nu}, A_{\alpha\nu} ] u_\mu \rangle+
i \lambda^{VA}_5 \langle [ \nabla_\alpha V^{\mu\nu}, A_{\mu\nu} ] u^\alpha \rangle \,,
\eqa
where $\langle...\rangle$ is short for the trace in flavor space and
as usual the chiral fields $u_\mu, f_{\pm}^{\mu\nu}, \chi_{\pm}, h_{\mu\nu}$
are defined in terms of the pseudo-Goldstone mesons and
external source fields \cite{rchpt89}. The SU(3) matrices
for vector and axial-vector resonances are given by
\bqa
V_{\mu\nu}= \left(
\begin{array}{ccc}
\frac{\rho_0}{\sqrt2}+\frac{\omega_8}{\sqrt6}+\frac{\omega_1}{\sqrt3} & \rho^+ & K^{*+} \nonumber \\
\rho^- & -\frac{\rho_0}{\sqrt2}+\frac{\omega_8}{\sqrt6}+\frac{\omega_1}{\sqrt3}  & K^{*0} \nonumber \\
K^{*-} & \overline{K}^{*0} &-\frac{2\omega_8}{\sqrt6}+ \frac{\omega_1}{\sqrt3} \nonumber \\
\end{array}
\right)_{\mu\nu} \,,
\eqa
\bqa
A_{\mu\nu}= \left(
\begin{array}{ccc}
\frac{a_1^0}{\sqrt2}+\frac{f_{1}^8}{\sqrt6}+\frac{f_{1}^1}{\sqrt3} & a_1^+ & K_{1A}^{+} \nonumber \\
a_1^- & -\frac{a_1^0}{\sqrt2}+\frac{f_{1}^8}{\sqrt6}+\frac{f_{1}^1}{\sqrt3}  & K_{1A}^{0} \nonumber \\
 K_{1A}^{-} & \overline{K}_{1A}^{0} & -\frac{2f_{1}^8}{\sqrt6}+\frac{f_{1}^1}{\sqrt3} \nonumber \\
 \end{array}
 \right)_{\mu\nu} \,,
\eqa and $K_{1A}$ is related to the physical states $K_{1}(1270), K_{1}(1400)$ through:
\bqa\label{theta}
K_{1A} &=& \cos\theta\,\, K_{1}(1400) + \sin\theta \,\,K_{1}(1270) \,.
\eqa
About the nature of $K_1(1270)$ and $K_{1}(1400)$, it has been proposed
in \cite{thetasuzuki} that they result from the mixing of $K_{1A}$ and $K_{1B}$,
where $K_{1A}$ denotes the strange partner of the axial vector resonance $a_1$
with $J^{PC}=1^{++}$ and $K_{1B}$ is the corresponding strange partner of
the axial vector resonance $b_1$ with $J^{PC}=1^{+-}$.
However in this paper, we will not include
the nonet of axial vector resonances with $J^{PC}=1^{+-}$. As argued in \cite{thetasuzuki},
the contributions from these kind of resonances to tau decays are proportional to the $SU(3)$
symmetry breaking effects. Moreover, as one can see later, we will assume the $SU(3)$ symmetry
for pseudo-Goldstone masses and also $SU(3)$ symmetry for both vector and axial
vector resonances in deriving the T-matrix throughout this article.
Physical masses will be taken into account in the kinematics.
For the vector resonances $\omega$ and $\phi$, we assume the ideal mixing for them
throughout this paper:
\bqa
\omega_1 = \sqrt{\frac{2}{3}} \omega - \sqrt{\frac{1}{3}} \phi \,, \nonumber \\
\omega_8 = \sqrt{\frac{2}{3}} \phi + \sqrt{\frac{1}{3}} \omega \,.
\eqa
In summary the R$\chi$T lagrangian that we will use in this paper is found to be
\be
\mathcal{L}_{R\chi T}= \mathcal{L}_2 +\mathcal{L}_{kin}(V, A)+\mathcal{L}_{2V, A}+
\mathcal{L}_{VVP}+\mathcal{L}_{VJP}+\mathcal{L}_{VAP}\,.
\ee

Using the R$\chi$T lagrangian above, we obtain the form
factors $v, a_1, a_2, a_3$ for $\tau^-(p) \rightarrow  K^-(p_1) \rho^0(p_2) \nu_\tau(q)$:
\bqa
&v =&-i\frac{ 2 F_V }{F_K M_\rho} [\,(d_1 + 8d_2)m_K^2 + d_3 (-m_K^2 +M_\rho^2 +s) \,]
 D_{K^*}(s)\nonumber\\ &&
+\,i\frac{ \sqrt2}{F_K M_V M_\rho} [\, (c_1 + c_2+8c_3-c_5)m_K^2 +
(c_2+c_5-c_1-2c_6)M_\rho^2+(c_1-c_2+c_5)s \,] \,,\nonumber \\ & &
\nonumber \\
& a_1 =& \frac{1}{4 M_\rho F_K} \bigg\{ \,\, F_V( m_K^2-M_\rho^2-s ) -2G_V( m_K^2+M_\rho^2-s )+
 \nonumber \\&&
2\sqrt2F_A  [\, \lambda_0 m_K^4  +
( M_\rho^2-s )( \lambda' M_\rho^2 -\lambda'' s ) -
m_K^2 (\lambda_0 M_\rho^2 +\lambda' M_\rho^2 +\lambda_0 s+\lambda'' s ) \,]
\nonumber \\ & &
\times [\,c_\theta^2  D_{K1H}(s)+ s_\theta^2  D_{K1L}(s)   \,] \,\,\bigg\} \,,
\nonumber \\ & &
\nonumber \\
&a_2 =& -\frac{ G_V M_\rho}{ F_K} D_{K}(s) +\frac{ \sqrt2 F_A M_\rho}{ F_K} (\lambda'+\lambda'')
[\,c_\theta^2  D_{K1H}(s)+ s_\theta^2  D_{K1L}(s)   \,]\,,
\nonumber \\ & &
\nonumber \\
&a_3 =&\frac{F_V- 2G_V}{2 F_K M_\rho} - \frac{ G_V M_\rho}{ F_K} D_{K}(s)+
\frac{  \sqrt2 F_A }{ F_K M_\rho} (\lambda_0 m_K^2 +\lambda'' M_\rho^2-\lambda'' s )
[\,c_\theta^2  D_{K1H}(s)+ s_\theta^2  D_{K1L}(s)   \,]\,,\nonumber \\
\eqa where $s=(p_1 + p_2)^2$ is the invariant mass of $\rho$ and
$K$; $c_\theta, s_\theta$ are short for $\cos\theta, \sin\theta$;
$F$ and $F_K$ are the decay constants for pion and kaon; $K,\rho,
K^*, K_{1L}, K_{1H}$ are used for kaon, $\rho(770)$, $K^*$(892),
$K_1$(1270) and $K_1$(1400) respectively. $D_K(s)$ denotes the
propagator for kaon \bqa\label{dpkaon} D_K(s) &=& \frac{1}{m_K^2 -
s}\, \eqa and the corresponding definition for the resonance $R$ is
\bqa\label{dpr} D_{R}(s) &=& \frac{1}{M_{R}^2 - s - i M_{R}
\Gamma_{R}(s)}\,, \eqa where the energy dependent decay width
$\Gamma_R(s)$ will be discussed in detail later. For convenience,
some useful combinations for $\lambda_i^{VA}$ have been used:
 \bqa
\sqrt2 \lambda_0 &=& -4\lambda^{VA}_1-\lambda^{VA}_2-
\frac{\lambda^{VA}_4}{2}-\lambda^{VA}_5 \,, \nonumber \\
\sqrt2 \lambda' &=& \lambda^{VA}_2-\lambda^{VA}_3+
\frac{\lambda^{VA}_4}{2}+\lambda^{VA}_5 \,, \nonumber \\
\sqrt2 \lambda''
&=&\lambda^{VA}_2-\frac{\lambda^{VA}_4}{2}-\lambda^{VA}_5 \,, \eqa
which have already been determined in \cite{vap}. In deriving the
expressions above, we have assumed $SU(3)$ symmetry for the
pseudoscalar masses in the T-matrix. Physical masses are taken into
account in the kinematics. The kaon decay constant $F_K$ have been
used in the form factors, instead of the parameter $F$ appearing in
the resonance chiral lagrangian. Likewise for the processes
containing pion in the final states, $F_\pi$ will be introduced. The
$\cO{(p^4)}$ corrections to $F_\pi, F_K$ in resonance chiral theory
have been studied in \cite{fkfpirxt}. Throughout this paper, instead
of stepping into the detail of the high order corrections to $F_\pi,
F_K$ in R$\chi$T, we will take the phenomenological values for them
in the numerical discussion. However when discussing the high energy
constraints, $SU(3)$ symmetry will be imposed to the
pseudo-Goldstone meson, i.e. only $F$ will enter the discussion of
QCD short distance constraints \cite{vap}. For the value of the
pseudo-Goldstone meson decay constant $F$ in $SU(3)$ limit , we will
use the pion decay constant to estimate it.  $F$ will be also used
to denote the pion decay constant in the remaining part of this
paper. The form factors for $\tau^- \rightarrow \omega K^-
\nu_\tau$, $\tau^- \rightarrow \phi K^-\nu_\tau$ and $\tau^-
\rightarrow \overline{K}^{*0} \pi^- \nu_\tau$ are very similar to
the ones of $\tau^- \rightarrow \rho^0 K^- \nu_\tau$. We give the
explicit expressions for these channels in the Appendix.

All of the processes above are driven by $\Delta S = 1$ currents ,
where both vector and axial-vector currents can take part in each
channel . In $\Delta S =0$ processes, such as $\tau^- \rightarrow
\rho^0 \pi^- \nu_\tau$, $\tau^- \rightarrow \omega \pi^- \nu_\tau$,
$\tau^- \rightarrow \phi \pi^- \nu_\tau$ and $\tau^- \rightarrow
K^*_0 K^- \nu_\tau$, not every channel can get contributions from
both vector and axial-vector currents. For  $\tau^- \rightarrow \rho^0
\pi^- \nu_\tau$, only the axial-vector current contributes, while
$\tau^- \rightarrow \omega \pi^- \nu_\tau$ is only driven by the
vector current. $\tau^- \rightarrow \phi \pi^- \nu_\tau$ vanishes in
our model, since it belongs to the next to leading order of $1/N_C$
expansion, which is beyond our scope. So in this paper we will not
discuss this process. The explicit expressions for the form factors
of $\Delta S = 0$ processes are also given in the Appendix.

Besides the lowest multiplet of resonances one can also introduce
heavier multiplets in R$\chi$T. However, including another multiplet,
 though trivial, may affect the well established relations of the
couplings for the lowest multiplet  \cite{rhoprime}.

The contribution to $\tau^- \rightarrow \rho^0 K^- \nu_\tau$ from
the new vector multiplet $V_1$ is \bqa \label{v1torhok} &&v^{V_1} =
-i\frac{ F_{V_1} }{F_K M_\rho} [\,d_m m_K^2 + d_M M_\rho^2 + d_s s
\,] D_{{K^*}'}(s) \,, \eqa where ${K^*}'$ corresponds to the
physical resonance $K^*$(1410) and we have used the following
lagrangian, given in \cite{rhoprime}, to get Eq.(\ref{v1torhok}):
\bqa
\mL_{2V_1}=\frac{F_{V_1}}{2\sqrt2}\langle {V_1}_{\mu\nu}
f_+^{\mu\nu}\rangle\,, \eqa
\bqa\label{v1lagr}
\mL_{V V_1 P}=&& d_a
\varepsilon_{\mu\nu\rho\sigma} \langle \{V^{\mu\nu},
V_1^{\rho\alpha}\} \nabla_{\alpha}u^{\sigma}  \rangle +
 d_b \varepsilon_{\mu\nu\rho\sigma} \langle \{V^{\mu\alpha}, V_1^{\rho\sigma}\}\nabla_{\alpha}u^{\nu}  \rangle +
\nonumber\\&&
d_c \, \varepsilon_{\mu\nu\rho\sigma} \langle \{ \nabla_\alpha V^{\mu\nu}, V_1^{\rho\alpha}\} u^{\sigma}  \rangle
+ d_d \varepsilon_{\mu\nu\rho\sigma} \langle
 \{ \nabla_\alpha V^{\mu\alpha}, V_1^{\rho\sigma}\} u^{\nu}  \rangle
\nonumber\\&&
+\, d_e \varepsilon_{\mu\nu\rho\sigma} \langle \{ \nabla^\sigma V^{\mu\nu}, V_1^{\rho\alpha}\} u_{\alpha}  \rangle
+\,i d_f \varepsilon_{\mu\nu\rho\sigma} \langle \{V^{\mu\nu}, V_1^{\rho\sigma}\} \chi_- \rangle \,.
\eqa
For the sake of simplicity, some combinations of $d_i$ from
Eq.(\ref{v1lagr}) have been defined in Eq.(\ref{v1torhok})
\bqa\label{dvd}
d_m &=&  d_a +d_b-d_c + 8d_f \,, \nonumber \\
d_M &=& d_b -d_a+ d_c -2d_d \,, \nonumber \\
d_s &=&d_c +d_a-d_b \,.
\eqa
The corresponding contributions from
the new vector multiplet to other channels will be included in the
form factors given in the Appendix.

To fulfill the QCD short-distance behavior \cite{lepage}, the vector
form factor $v$ defined in Eq.(\ref{general}) should vanish at high
energy limit, which gives us a new constraint to the resonance
couplings
\bqa\label{newc65} c_6-c_5 = \frac{2 d_3 F_V + d_s
F_{V_1}}{2\sqrt2 M_V}\,,
\eqa
and also allows us to recover the constraint already given in \cite{vvp}:
\bqa\label{c125}
c_1-c_2+c_5=0\,.
\eqa
Both the chiral symmetry and $SU(3)$ symmetry for vector resonances have
been assumed in deriving the above constraints. Another constraint from
\cite{vvp} that will be  useful for us is
\bqa\label{c13}
c_1+4c_3=0\,,
\eqa
which, like the constraint in Eq.(\ref{c125}), will not be influenced
by including the new multiplet \cite{rhoprime}.
For the axial-vector current, since we do not include extra operators
for axial-vector resonances in this paper, the constraints on
the axial-vector resonance couplings are the same as those in \cite{vap}.

\section{Phenomenological discussion}

To perform the numerical discussion,  the values of related
parameters will be taken from \cite{vap}: \bqa F_V^2 = F^2
\frac{M_A^2}{M_A^2-M_V^2},\qquad F_A^2 = F^2
\frac{M_V^2}{M_A^2-M_V^2},\qquad G_V^2 =
F^2\frac{M_A^2-M_V^2}{M_A^2}, \eqa where $M_V$ and $M_A$ denote the
masses for vector and axial-vector resonances in large-$N_C$ limit.
As pointed in \cite{rhoprime}, $M_V$ can be safely estimated by the
mass of $\rho(770)$ meson, while for the axial-vector resonance,
$M_A$ is apparently different from the mass of $a_1(1260)$ and
$M_A=0.998(49)$GeV is obtained using the experiment value for the
axial-vector form factor in \cite{rhoprime}. Throughout this paper,
we will use $M_V=M_\rho$ and $M_A = 0.998$GeV to evaluate resonance
couplings, but masses appearing in the kinematics will take their
values from \cite{pdg}. The explicit values for resonance couplings
that we use in the fit are given by \be\label{fvnc}
 F = 0.0924,\quad F_K = 0.113,\quad F_V = 0.147, \quad F_A = 0.114,\quad G_V =0.058\,,
\ee in units of GeV and the values of the pion decay constant $F$
and kaon decay constant $F_K$ are taken from \cite{fkexp}. Using the
decay widths of $\rho \rightarrow e^+ e^-$, $a_1 \rightarrow
\pi\gamma$ and $\rho \rightarrow \pi\pi$, one can respectively
estimate  the values for $F_V, F_A, G_V$: \be F_V = 0.156,\qquad F_A
= 0.122,\qquad G_V =0.066, \ee which are given in units of GeV. One
can see that they are reasonably consistent with the theoretical
determinations given in Eq.(\ref{fvnc}).

The resonance couplings  $\lambda'$, $\lambda''$ and $\lambda_0$,
which are related to axial-vector resonances, have also been given
in terms of the masses of resonances in \cite{vap}
\bqa
&& \lambda'= \frac{M_A}{2\sqrt2 M_V}, \qquad
\lambda'' = \frac{M_A^2-2M_V^2}{2\sqrt2 M_A M_V}, \qquad
\lambda_0 =\frac{\lambda'+\lambda''}{4} \,.
\eqa
One can easily get the values:
\be
\lambda'=0.455,\qquad \lambda''=-0.0938,\qquad \lambda_0=0.0904\,.
\ee
However in later discussion, to test the stability,
we will also perform our fit using another set of values for $\lambda_i$:
\be
\lambda' =0.5, \qquad \lambda'' = 0, \qquad\lambda_0 =0.125,
\ee
which can be derived by assuming the original KSRF relation $F_V=2G_V$ \cite{vap}.

For the heavier vector multiplet parameter $F_{V_1}$, we determine
its value using the decay widths of $V_1 \rightarrow e^+e^-$, which
will be discussed in detail in the following subsection. Other
parameters related to the heavier vector multiplet $d_m, d_M,
d_s$, will be fit later. Due to the inclusion of the new vector
multiplet, some of the well established constraints in \cite{vvp},
such as $d_1+8d_2,\, d_3$, will get corrections \cite{rhoprime}.
However, since the combination of $d_1+8d_2$ is the coefficient of
the mass of pseudo-Goldstone mesons, the final results should be rather
insensitive to the value of this combination. So we will still take
the value from \cite{vvp} as an approximation. While for $d_3$, we
will fit its value in the  $\tau^- \rightarrow \omega \pi^-
\nu_\tau$ spectral function. The mixing angle $\theta$ in
Eq.(\ref{theta}) will  be fit in the invariant
 mass distribution of $\omega K$ system from $\tau^- \rightarrow \omega K^- \nu_\tau$.
So in total, we have 5 free parameters to fit: $d_3$, which is a
resonance coupling related to the lowest vector multiplet;
 $d_m, d_M, d_s$, which are the couplings for the
excited vector multiplet $V_1$; the mixing angle $\theta$, which is
a parameter for the axial-vector resonances defined in
Eq.(\ref{theta}). For the remaining parameters, which have not been
mentioned above, we will take their values from \cite{pdg}.

\subsection {Determination of the parameter $F_{V_1}$ for the heavier vector multiplet}

Since we have included a set of new multiplet for vector resonances,
a corresponding set of physical states has to be assigned to this
multiplet and a natural choice from the particle lists presented in
 \cite{pdg} should be
\bqa\label{v1mem} \left\{ \rho(1450), \omega(1420), \phi(1680),
K^*(1410) \right\} \in V_1 \,.
\eqa
As for $\omega$ and $\phi$, ideal mixing will also
be assumed for $\omega'$ and $\phi'$.

For $F_{V_1}$  we use the decay widths of $V_1\rightarrow e^+ e^-$ to
determine its value. The decay width for a vector resonance
 $V \rightarrow e^+ e^-$ can be deduced using
$\frac{F_V}{2\sqrt2}\langle V_{\mu\nu} f_+^{\mu\nu}\rangle$. The
expression for the decay width of $V \rightarrow e^+ e^-$ is found to be
\bqa
 \Gamma_V^{e^+e^-} = \frac{1}{3}
\frac{1}{8\pi}\,\frac{\sqrt{M_V^2-4m_e^2}}{2M_V^2}\,\, \frac{64
\alpha^2 \pi^2 F_V^2 ( 2m_e^2 + M_V^2)}{M_V^2} \,,
\eqa
where $\alpha$ is the fine structure constant;  $V$ is a vector resonance
with isospin $I=1$. For $I=0$ states, such as $\omega$ and $\phi$ like
vector resonances, one has to multiply the above formula with
$\frac{1}{9}$ and $\frac{2}{9}$ respectively due to their different
couplings with the photon.

The available decay widths of the vector resonances in the new
multiplet to $e^+ e^-$ can be extracted from \cite{pdg}:
$\Gamma_{\rho(1450)} \sim 1.8$ KeV, $\Gamma_{\omega(1420)} \sim 0.12$ KeV,
where we have used
$\Gamma_{\pi\pi}\times\Gamma_{e^+e^-}/\Gamma_{\textrm{total}}= 0.12$ KeV,
$\Gamma_{\pi\pi}/\Gamma_{\omega\pi}=0.32$ ,
$\Gamma_{\omega\pi}/\Gamma_{\textrm{total}}= 0.21$ to get
$\Gamma_{\rho(1450)\rightarrow e^+e^-}$ and
$\Gamma_{\rho\pi}\times\Gamma_{e^+e^-}/\Gamma_{\textrm{total}}= 0.081$KeV,
$\Gamma_{\rho\pi}/\Gamma_{\textrm{total}}=0.7$
to extract the value for $\Gamma_{\omega(1420)\rightarrow e^+e^-}$.
Using this set of data allows us to make an estimate for $F_{V_1}$:
\bqa
\Gamma_{\rho(1450)}= 1.8 \, \textrm{KeV} \,\,\, \Rightarrow \,\,\, |F_{V_1}| = 0.11 \, \textrm{GeV} \,, \nonumber \\
\Gamma_{\omega(1420)}= 0.12\, \textrm{KeV} \,\,\,\Rightarrow \,\,\,
|F_{V_1}| = 0.08 \, \textrm{GeV} \,, \eqa where one can see that the
assignment (\ref{v1mem}) seems reasonable. Although due to the poor
knowledge for $\rho(1450)$ and $\omega(1420)$, different sets of
values for $\Gamma_{\rho(1450)\rightarrow e^+e^-}$ and
$\Gamma_{\omega(1420)\rightarrow e^+e^-}$ can be extracted from
different experimental groups results \cite{pdg}, most of them lead
to a prediction for $|F_{V_1}|$ around 0.1GeV. About the sign of
$F_{V_1}$, since in our case what appears in the T-matrix is always
the combination of $F_{V_1} (\,d_m  m_\pi^2 + d_M  M_V^2 + d_s s )$,
this allows us to fix the sign of $F_{V_1}$ and to leave $d_m, d_M,
d_s$ free. So in later discussion, we will set $F_{V_1}= -0.1$GeV.
To determine the free parameters $d_m, d_M, d_s$, which are related
to the new vector multiplet, we will fit the $\tau^- \rightarrow
\omega\pi^-\nu_\tau$  spectral function and the invariant mass
distribution for $\omega K$ in the decay of $\tau^- \rightarrow
\omega K^- \nu_\tau$.

\subsection{Introduce the energy dependent decay widths for intermediate resonances}
Before stepping into the fit, some points about the decay widths of
the intermediate resonances appearing in $\tau$ decays will be
stressed. Since most of the intermediate resonances have wide decay
widths, the off-shell widths of these resonances may play an
important role in the dynamics of $\tau$ decays. To introduce the
finite decay widths for the resonances implies that the corrections
from the next-to-leading order of $1/N_C$ expansion are taken
account into our game. This issue has been discussed in
\cite{engwidthrho} for the decay width of $\rho(770)$ and we take
the result of that article \be\label{gammarho} \Gamma_\rho(s) =
\frac{s M_V}{96 \pi F^2}\left[ \sigma_\pi^3 \theta(s-4m_\pi^2) +
\frac{1}{2}\sigma_K^3 \theta(s-4m_K^2) \right]\,, \ee where
$\sigma_P =\sqrt{1-4m_P^2/s}$ and $\theta(s)$ is the step function.
About the energy dependent widths for $\rho', K^*, {K^*}',
K_1(1270), K_1(1400), a_1(1260)$, we follow the similar way
introduced in\cite{taupik}\cite{engwidthrhopkv} to construct them:
\bqa
 \Gamma_{\rho'}(s) &=& \Gamma_{\rho'}
\frac{s}{M_{\rho'}^2}\left[\frac{ \sigma_{\pi\pi}^3(s)  +
\frac{1}{2}\sigma_{K K}^3 }
{\sigma_{\pi\pi}^3(M_{\rho'}^2)  + \frac{1}{2}\sigma_{KK}^3(M_{\rho'}^2) } \right] \,,\nonumber \\
\Gamma_{K^*}(s) &=& \Gamma_{K^*}
\frac{s}{M_{K^*}^2}\left[\frac{ \sigma_{K\pi}^3(s)  + \sigma_{K\eta}^3(s)}
{\sigma_{K\pi}^3(M_{K^*}^2)  + \sigma_{K\eta}^3(M_{K^*}^2)} \right] \,,
\nonumber \\
\Gamma_{{K^*}'}(s) &=&\Gamma_{{K^*}'}
\frac{s}{M_{{K^*}'}^2}\left[\frac{ \sigma_{K\pi}^3(s)  + \sigma_{K\eta}^3(s)}
{\sigma_{K\pi}^3(M_{{K^*}'}^2)  + \sigma_{K\eta}^3(M_{{K^*}'}^2)} \right] \,,
\nonumber \\
\Gamma_{K_{1L}}(s) &=&\Gamma_{K_{1L}}
\frac{s}{M_{K_{1L}}^2}\left[\frac{ \sigma_{K\rho}^3(s)  + \sigma_{K^*\pi}^3(s)}
{\sigma_{K\rho}^3(M_{K_{1L}}^2)  + \sigma_{K^*\pi}^3(M_{K_{1L}}^2)} \right] \,,
\nonumber \\
\Gamma_{K_{1H}}(s) &=& \Gamma_{K_{1H}}
\frac{s}{M_{K_{1H}}^2}\left[\frac{ \sigma_{K\rho}^3(s)  + \sigma_{K^*\pi}^3(s)}
{\sigma_{K\rho}^3(M_{K_{1H}}^2)  + \sigma_{K^*\pi}^3(M_{K_{1H}}^2)} \right] \,,
\nonumber \\
\Gamma_{a_{1}}(s) &=&\Gamma_{a_{1}}
\frac{s}{M_{a_{1}}^2}\left[\frac{ \sigma_{\pi\rho}^3(s)  +
\frac{1}{2}\sigma_{K K^*}^3(s)} {\sigma_{\pi\rho}^3(M_{a_{1}}^2)  +
\frac{1}{2}\sigma_{K K^*}^3(M_{a_{1}}^2)} \right] \,,
\eqa
where
\be
\sigma_{P Q}(s) = \frac{1}{s}\sqrt{\left(s-(m_P+m_Q)^2
\right)\left(s-(m_P-m_Q)^2\right)}\,\theta\left( s-(m_P+m_Q)^2
\right)\,;
\ee
and the constant parameter $\Gamma_R$ appearing in
the energy dependent width will be fixed at the central value of the
corresponding resonance decay width given in \cite{pdg}.
For $a_1(1260)$, $\Gamma_{a_1}=0.5$ GeV will be taken.
In the following sections, the energy dependent widths
will be always implemented into our discussion.

\subsection{Predictions for branching ratios only including the lowest multiplet}

In this subsection,  we perform our discussion only including the
lowest multiplet resonances for $\tau$ decays. If the values for
vector resonance couplings are taken from \cite{vvp}\cite{vap} and
for axial-vector resonance couplings we take $\lambda' =
\frac{1}{2}, \, \lambda'' =0, \, \lambda_0 =
\frac{\lambda'+\lambda''}{4}, \theta =45^{\textrm{o}}$ (assuming ideal
mixing), the theoretical predictions for branching ratios are
summarized in Table \ref{tab1} for $\Delta S =0$ process and Table \ref{tab2}
for $\Delta S =1$ process, which are presented in the next subsection.

We can see that the largest gap between the theoretical prediction
and experiment data happens in $\tau^-\rightarrow
\omega\pi^-\nu_\tau$ channel. Since only vector current enters into
$\tau^-\rightarrow \omega\pi^-\nu_\tau$, it implies that only
including the lowest vector resonance in this channel is not enough
and $d_3$ determined in \cite{vvp} will get non-negligible
corrections when extra multiplet is introduced. Therefore it can be
a perfect channel to investigate vector resonances. Moreover $\tau^-
\rightarrow \rho^0\pi^-\nu_\tau$ can be an excellent process to
study the axial-vector resonance, since only the axial-vector current
takes part in this channel. However the data for this channel is
absent.

In the following discussion, we will always include two multiplets
for vector resonances and fit unknown couplings in the
$\tau^-\rightarrow \omega\pi^- \nu_\tau$ spectral function and the
invariant mass distribution for $\omega K$ in $\tau^-\rightarrow
\omega K^- \nu_\tau$.

\subsection{Fitting results }
The experiment data of the $\tau^-\rightarrow \omega\pi^- \nu_\tau$
spectral function is given in \cite{expwpi} and the invariant mass
distribution of $\omega K$ system in $\tau^-\rightarrow \omega K^-
\nu_\tau$ can be found in \cite{expwk}.

Following the definition of spectral functions for $\tau$ decays
first given in \cite{defspec} and recently summarized in
\cite{taurevmp}, the explicit expression of the $\tau^-\rightarrow
\omega\pi^- \nu_\tau$ spectral function in our model can be written as
\bqa
V(s) =&& \frac{1}{6 F^2 M_\omega^2 \pi s^2 S_{EW}}\left[
m_\pi^4+(M_\omega^2-s)^2-2m_\pi^2(M_\omega^2+s)\right]^{3/2}
\nonumber\\ && \times \bigg|( 2d_3 F_V + d_s
F_{V_1})\frac{M_\omega^2}{M_V^2}+ F_{V_1}(d_m m_\pi^2 + d_M
M_\omega^2 +d_s s ) D_{\rho'}(s)  \nonumber \\&& + 2 F_V [(d_1+8 d_2
)m_\pi^2 + d_3(s + M_\omega^2 -m_\pi^2)] D_{\rho}(s) \bigg|^2\,,
\eqa
where Eq.(\ref{newc65}), Eq.(\ref{c125}) and Eq.(\ref{c13})
have been used and the value of the electroweak correction
factor $S_{EW}$, which has been analyzed in \cite{sew}, will
be taken as $S_{EW}=1.0194$ (at the scale $m_\tau$).

In the fit of the $\tau^-\rightarrow \omega\pi^- \nu_\tau$ spectral
function, we set $F_{V_1} = -0.1 \textrm{GeV}, d_m = -1.0$ and fit
the parameters $d_3, d_M, d_s$. The fitting results are
\bqa\label{fitwpi131}
d_3 &=& -0.25 \pm 0.01, \nonumber \\
d_M &=& 0.99 \pm 0.08, \nonumber \\
d_s &=& -0.29 \pm 0.03, \eqa with $\chi^2/d.o.f = 25.8/13\simeq 2.0$
.

The decay width for $\rho'$ given in \cite{pdg} is $0.4\pm 0.06$
GeV. The fitting results above are based on $\Gamma_{\rho'}=0.4$ GeV
. If we take $\Gamma_{\rho'}=0.34$ GeV, we find the $\chi^2$ in the
fit will get better. The fitting results then are
\bqa\label{fitwpi132}
d_3 &=& -0.25 \pm 0.01, \nonumber \\
d_M &=& 0.97 \pm 0.08, \nonumber \\
d_s &=& -0.27 \pm 0.03, \eqa with $\chi^2/d.o.f = 21.8/13\simeq 1.7$
.

In the fit of the invariant mass distribution of
$\omega K$, we take Eq.(\ref{fitwpi132}) as our inputs.
So we  set $d_m=-1.0, d_s=-0.27, d_3=-0.25$ and take different values of
 $\lambda', \lambda'', \lambda_0$ to fit $d_M, \theta$ :
\begin{itemize}
\item
In case of $\lambda' = \frac{M_A}{2\sqrt2 M_V}, \, \lambda'' =
\frac{M_A^2-2M_V^2}{2\sqrt2}, \, \lambda_0 =
\frac{\lambda'+\lambda''}{4}$:

 The results are
\bqa\label{fitwk131}
d_M &=& 0.64 \pm 0.93, \nonumber \\
\cos^2\theta &=& 0.26 \pm 0.11,
\eqa
with $\chi^2/d.o.f \simeq 3.4/5=0.68$ .

\item
In case of $\lambda' = 0.5, \, \lambda'' =0, \, \lambda_0 =0.125$:

  The results are
\bqa\label{fitwk132}
d_M &=& 0.64 \pm 0.87, \nonumber \\
\cos^2\theta &=& 0.28 \pm 0.12,
\eqa
with $\chi^2/d.o.f \simeq 3.3/5 = 0.66$ .

\end{itemize}

Since what we can fit in the two processes for the couplings $d_m,
d_M$  are actually the combinations of $(d_m m_\pi^2 + d_M
M_\omega^2)$ in $\tau^-\rightarrow \omega \pi^- \nu_\tau$ and  $(d_m
m_K^2 + d_M M_\omega^2)$ in $\tau^-\rightarrow \omega K^- \nu_\tau$,
the true values for $d_m, d_M$ can be solved using the values we
have got in the two processes :
\begin{itemize}
\item
For $\lambda' = \frac{M_A}{2\sqrt2 M_V}, \, \lambda'' =
\frac{M_A^2-2M_V^2}{2\sqrt2}, \, \lambda_0 =
\frac{\lambda'+\lambda''}{4}$:

\be
d_m= -1.90,\qquad d_M= 1.00 \,\,\,.
\ee

\item For $\lambda' =0.5, \, \lambda'' =0, \, \lambda_0 =0.125$:
\be
d_m= -1.90,\qquad d_M= 1.00 \,\,\,.
\ee

\end{itemize}

With the above values of resonance couplings, the predictions for
branching ratios we get are summarized in Table \ref{tab1} and Table \ref{tab2}.

\begin{table}[h]
\begin{tabular}{|c|c|c|c|c|}
\hline
 &  Exp & One multiplet & Fit 1 & Fit 2 \\
\hline
$B(\tau^-\rightarrow \rho^0\pi^-\nu_\tau)$\,&\,--- \,&\,  $8.1 \times 10^{-2}$\,
&\,$9.4 \times 10^{-2}$ \,&\,$8.1 \times 10^{-2}$\\
\hline
 $B(\tau^-\rightarrow \omega\pi^-\nu_\tau)$ \,&\, $(1.95 \pm 0.08) \times 10^{-2}$ \,
 &\, $0.17  \times 10^{-2}$\,&\, $ 2.1
\times 10^{-2}$ \,&\, $2.1  \times 10^{-2}$\\
\hline
$B(\tau^-\rightarrow K^{*0}K^-\nu_\tau)$\,&\, $(2.1 \pm 0.4) \times 10^{-3}$ \,
&\, $1.4 \times 10^{-3} $\,&\, $1.5 \times 10^{-3}$ \,&
\,$1.5 \times 10^{-3}$ \\
\hline
\end{tabular}
\caption{Branching ratios for $\Delta S =0$ processes.
The second column denotes experimental values, which are taken from \cite{pdg}.
The values from the third column to the fifth column denote our predictions
under different assumptions: only including the lowest
multiplet, the fitting results with
$\lambda' = \frac{M_A}{2\sqrt2 M_V}, \, \lambda'' = \frac{M_A^2-2M_V^2}{2\sqrt2}, \,
\lambda_0 = \frac{\lambda'+\lambda''}{4}$ (Fit 1) and
the fitting results with $\lambda' =0.5, \, \lambda'' =0, \, \lambda_0 = 0.125$ (Fit 2)\,. \label{tab1}}
\end{table}

\begin{table}[h]
\begin{tabular}{|c|c|c|c|c|}
\hline
  & Exp & One multiplet & Fit 1 & Fit 2 \\
\hline
$B(\tau^-\rightarrow \rho^0K^-\nu_\tau)$\,&\, $(1.6 \pm 0.6)  \times 10^{-3} $  \,
&\,  $3.9  \times 10^{-4} $ \,& \, $4.7  \times 10^{-4} $ \,&\, $3.5  \times 10^{-4}$ \\
\hline
$B(\tau^-\rightarrow \omega K^-\nu_\tau)$\,&\,$(4.1 \pm 0.9)   \times 10^{-4}$  \,
&\, $3.5  \times 10^{-4}$ \,&
\,$ 4.0  \times 10^{-4}$ \,&\, $3.0  \times 10^{-4}$ \\
\hline
$B(\tau^-\rightarrow \phi K^-\nu_\tau)$\,&\, {\footnotesize $(4.05 \pm 0.25 \pm 0.26)  \times 10^{-5}$(Belle)}\,
&\, $1.7 \times 10^{-5}$ \, &\,$1.8 \times 10^{-5}$ \,&\, $1.6 \times 10^{-5}$ \\
\,&\, {\footnotesize $(3.39 \pm 0.20 \pm 0.28)\times 10^{-5}$(BaBar)} \,&\,  \,&\, \,&\,\\
\hline
$B(\tau^-\rightarrow \overline{K}^{*0}\pi^-\nu_\tau)$\,&\, $(2.2 \pm 0.5) \times 10^{-3}$  \,&
\, $3.3  \times 10^{-3}$ \,&\,$5.1  \times 10^{-3}$ \,&\, $4.0 \times 10^{-3}$ \\
\hline
\end{tabular}
\caption{Branching ratios for $\Delta S =1$ processes.
The meaning of numbers in different columns is the same to Table \ref{tab1}.
The experimental values for $\phi K^-$ channel are taken from
\cite{expphikbelle} and \cite{expphikbabar}.
The remaining experimental data is taken from \cite{pdg}. \label{tab2}}

\end{table}

At the end of this section, some comments about the fitting results are given below:
\begin{enumerate}
\item For $\rho K^-$, $\omega K^-$ and $\overline{K}^{*0}\pi^-$ channels:
The branching fractions for $\tau^- \rightarrow \omega K^-\nu_\tau$
with different choices for $\lambda_i$ are both consistent with the
experimental value. However in our model we cannot explain the issue
on the small ratio of the branching fraction
$\frac{B(\tau^-\rightarrow \omega K^-\nu_\tau)}{B(\tau^-\rightarrow
\rho K^-\nu_\tau)}$ raised in \cite{expwk}, since in our case the
dynamics for the two processes are the same and the only difference
is the kinematics, which is very tiny in this case. Comparing the
experimental data for $\overline{K}^{*0}\pi^-$ channel, our
predictions seem larger. However taking into account that we  work in
the leading order of $1/N_C$ expansion and take the $SU(3)$ symmetry for
vector and axial-vector resonances in the T-matrix, the prediction for
the $\overline{K}^{*0}\pi^-$ process can be acceptable.

\item For $\phi K^-$ channel: As we will see later, although
our prediction for the invariant mass distribution of $\phi K$ in
the decay of $\tau^- \rightarrow \phi K^- \nu_\tau$ seems reasonably
consistent with the experimental data, the prediction for the
branching ratio is around $50\%$ of the experimental value.

\item  For $\Delta S =0$ channel:
Since we fit the spectral function for $\omega \pi$ process, the
branching ratio for $\tau^-\rightarrow \omega \pi^-\nu_\tau$ is
always perfect. The corresponding prediction for $\tau^-\rightarrow
K^{*0} K^-\nu_\tau$ is also reasonable comparing with the
experimental data and it is insensitive to the values of
axial-vector resonance couplings $\lambda_i$. Although the branching
ratio for $\tau^- \rightarrow \rho^0\pi^-\nu_\tau$ is absent in
\cite{pdg}, the branching fraction of $\tau^- \rightarrow
\pi^-\pi^+\pi^-\nu_\tau$ can be a reference for the $\tau^- \rightarrow
\rho^0\pi^-\nu_\tau$ process. Comparing the experimental
data \mbox{$B(\tau^- \rightarrow \pi^-\pi^+\pi^-\nu_\tau(\textrm{ex}.
K^0,\omega))=(8.99\pm 0.08)\times 10^{-2}$} \cite{pdg}, our
predictions indicate that the branching ratio of $\tau^- \rightarrow
\pi^-\pi^+\pi^-\nu_\tau$ is dominantly contributed by the $\tau^-
\rightarrow \rho^0\pi^-\nu_\tau$ channel.

\item In the fit of the invariant mass distribution of $\omega K^-$,
we get a rather large error for the coupling $d_M$. To understand
this, we have made a detail analysis for this process. If we switch
off all of the other contributions except ${K^*}'$ resonance in
$\tau^- \rightarrow \omega K^- \nu_\tau$, the remaining branching
ratio is only around $12\%$ of the total. Hence the large error we
get for $d_M$, a resonance coupling related with ${K^*}'$, is not
hard to understand. If the contributions from $K_1(1270)$ and
$K_1(1400)$ are turned off, we find the remaining parts  contribute
around $38\%$ of the total branching ratio and the situations in
$\rho K^-$ and $\overline{K}^{*0} \pi^-$ channels are similar to
$\omega K^-$ channel, which indicates that the axial-vector
resonances play a rather important role in these processes. However
this phenomenon does not happen in $\phi K^-$ channel. In contrast,
if we switch off all of the other contributions except $a_1$
resonance in $\tau^- \rightarrow K^{*0}K^-\nu_\tau$ channel, which
is driven by the $\Delta S =0$ current, only $23\%$ of the total
branching ratio can be reached.

\item In case of $\lambda'=0.5, \lambda''=0, \lambda_0=0.125$, our
fitting result for $\theta$ is $\cos^2\theta = 0.28 \pm 0.12$,
which indicates $|\theta|=58.1^{+8.4}_{-7.3}$ degrees. The result in
the other case is quite similar. Our result for $\theta$ is
consistent with $|\theta| = 37^{\textrm{o}}$ and $58^{\textrm{o}}$
recently determined also in $\tau$ decays \cite{thetahycheng}.
Our fit favors the larger value.

\item About the uncertainties of our predictions for the branching ratios,
taking into account the approximation we have made throughout this paper,
such as working in the leading order of $1/N_C$ expansion and
$SU(3)$ symmetry for vector and axial-vector resonances in the T-matrix,
and also considering the large error of the resonance couplings that we get
from the fit of $\tau^-\rightarrow \omega K^-\nu_\tau$, a conservative
estimate of the uncertainties of our predictions for the branching
ratios should be around thirty percent.

\item Comparison of the figures we have obtained for the
$\tau^- \rightarrow \omega \pi^- \nu_\tau$ spectral function and the
invariant mass distributions for $\omega K^-, \phi K^-$ between the
experimental data are given in Fig.(\ref{fig1}-\ref{fig3}) respectively.
Although different choices for $\lambda_i$ affect the branching ratios, the
invariant mass distributions are barely influenced. So we only plot
the figures with
 $\lambda'=0.5, \lambda''=0, \lambda_0=0.125$.
\vspace{0.2cm}
\begin{figure}[ht]
\begin{center}
\includegraphics[angle=0, width=0.55\textwidth]{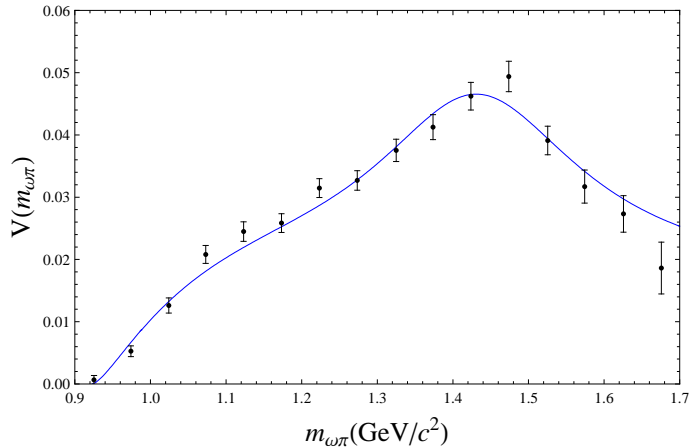}
\caption{Spectral function for $\tau^-\rightarrow\omega\pi^-\nu_\tau$. The experimental data are taken from \cite{expwpi}.
\label{fig1} }
\end{center}
\end{figure}

\begin{figure}[ht]
\begin{center}
\includegraphics[angle=0, width=0.55\textwidth]{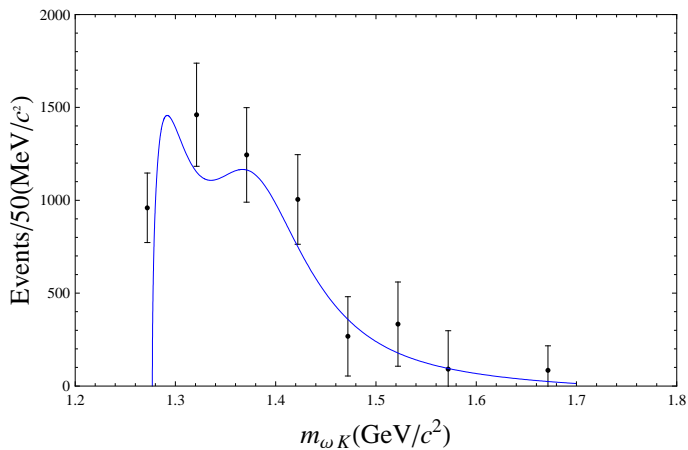}
\caption{Invariant mass distribution for $\omega K^-$ in the process of $\tau^-\rightarrow\omega K^-\nu_\tau$.
The experimental data are taken from \cite{expwk}. \label{fig2}}
\end{center}
\end{figure}

\begin{figure}[ht]
\begin{center}
\includegraphics[angle=0, width=0.55\textwidth]{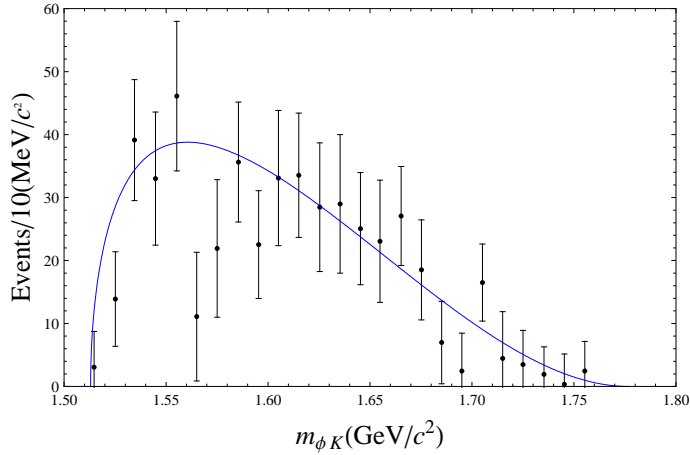}
\caption{Invariant mass distribution for $\phi K^-$ in the process of $\tau^-\rightarrow\phi K^-\nu_\tau$.
 The experimental data are taken from \cite{expphikbelle}, where only the data up to
 $m_{\phi K}=1.75$ GeV are quoted in the plot.\label{fig3}}
\end{center}
\end{figure}

\end{enumerate}

\section{Conclusion}
In this work, the resonance chiral effective theory is exploited to
study a $\tau$ decaying into a vector resonance plus a
pseudo-Goldstone meson and a $\tau$ neutrino. Two multiplets of
vector resonances have been introduced in our discussion. We fit the
$\tau^- \rightarrow \omega \pi^- \nu_\tau$ spectral function  and
decay distribution of $\tau^- \rightarrow \omega K^- \nu_\tau$ to
get unknown resonance couplings. Then we make a prediction for
branching ratios for all channels. Taking into account the
approximation we have made in this paper, like taking the leading
order of $1/N_C$ expansion and $SU(3)$ symmetry for vector and
axial-vector resonances in the T-matrix, we conclude that resonance
chiral effective theory can describe the experimental data
reasonably, although  the issue of the small ratio
$\frac{B(\tau^-\rightarrow \omega K^-\nu_\tau)}{B(\tau^-\rightarrow
\rho K^-\nu_\tau)}$ raised in \cite{expwk} is still there. Our
fit for the mixing angle $\theta$, which is defined in
Eq.(\ref{theta}) to describe the mixture between the flavor
eigenstates of the axial-vector resonances and physical
states $K_1(1270), K_1(1400)$, leads to $|\theta|\simeq 58^{\textrm{o}}$, which is consistent
with previous determination with $|\theta|=37^{\textrm{o}}$ and $58^{\textrm{o}}$ also from
tau decays \cite{thetahycheng}.

\section*{ACKNOWLEDGEMENTS}
I really appreciate the help of Jorge Portol\'es for enlightening me on this
subject. I am greatly indebted to him for helpful discussions and also valuable
suggestions on the manuscript. I also would like to thank Han-Qing Zheng
for reading the manuscript and precious suggestions.
Z.H.G is funded in part by China Scholarship Council and
National Nature Science Foundation of China under grant number
10721063 and 10575002. This work is also partially supported by EU Contract
No. MRTN-CT-2006-035482 (FLAVIAnet), by MEC (Spain) under
grant FPA2007-60323 and by Spanish Consolider-Ingenio 2010 Programme
CPAN (CSD2007-00042).

\section*{APPENDIX: Explicit expressions of form factors for  other channels}
\setcounter{equation}{0}
\renewcommand{\theequation}{A-\arabic{equation}}
The form factors of $v, a_1, a_2, a_3$ for other channels derived by
$\Delta S =1$ currents are given by
\begin{itemize}
\item $\tau^-(p) \rightarrow  K^-(p_1) \omega(p_2) \nu_\tau(q)$:
\bqa
&v =&-i\frac{ F_{V_1} }{F_K M_\omega} [\,d_m m_K^2 + d_M
M_\omega^2 + d_s s \,] D_{{K^*}'}(s) \nonumber\\ && - i\frac{ 2 F_V
}{F_K M_\omega} [\,(d_1 + 8d_2)m_K^2 + d_3 (-m_K^2 +M_\omega^2 +s)
\,] D_{K^*}(s)\nonumber\\ && +\,i\frac{ \sqrt2}{F_K M_V M_\omega}
[\, (c_1 + c_2+8c_3-c_5)m_K^2 +
(c_2+c_5-c_1-2c_6)M_\omega^2+(c_1-c_2+c_5)s \,] \,, \nonumber \\&&
\nonumber \\
&a_1 =& \frac{1}{4 M_\omega F_K} \bigg\{\,\, F_V( m_K^2-M_\omega^2-s )
-2G_V( m_K^2+M_\omega^2-s )+ \nonumber \\ && +
 2\sqrt2 F_A [\, \lambda_0 m_K^4  +
( M_\omega^2-s )( \lambda' M_\omega^2 -\lambda'' s ) -
m_K^2 (\lambda_0 M_\omega^2 +\lambda' M_\omega^2 +\lambda_0 s+\lambda'' s ) \,] \nonumber \\ &&
\times [\,c_\theta^2  D_{K1H}(s)+ s_\theta^2  D_{K1L}(s)   \,]     \,\,\bigg\}\,,
\nonumber \\&&
\nonumber \\
&a_2 =& -\frac{ G_V M_\omega}{ F_K} D_{K}(s)
+\frac{ \sqrt2 F_A M_\omega}{ F_K} (\lambda'+\lambda'')
[\,c_\theta^2  D_{K1H}(s)+ s_\theta^2  D_{K1L}(s)   \,]\,,
\nonumber \\&&
\nonumber \\
&a_3 =&\frac{F_V- 2G_V}{2 F_K M_\omega} - \frac{ G_V M_\omega}{ F_K} D_{K}(s)+\nonumber \\ &&
 +\frac{  \sqrt2 F_A }{ F_K M_\omega} (\lambda_0 m_K^2 +\lambda'' M_\omega^2-\lambda'' s )
[\,c_\theta^2  D_{K1H}(s)+ s_\theta^2  D_{K1L}(s)   \,]\,.
\eqa

\item $\tau^-(p) \rightarrow  K^-(p_1) \phi(p_2) \nu_\tau(q)$:
\bqa
&v =&i\frac{\sqrt2 F_{V_1} }{F_K M_\phi} [\,d_m m_K^2 + d_M M_\phi^2 + d_s s \,] D_{{K^*}'}(s)
\nonumber\\ &&
+ i\frac{ 2\sqrt2 F_V }{F_K M_\phi} [\,(d_1 + 8d_2)m_K^2
+ d_3 (-m_K^2 +M_\phi^2 +s) \,] D_{K^*}(s)\nonumber\\ &&
-\,i\frac{ 2}{F_K M_V M_\phi} [\, (c_1 + c_2+8c_3-c_5)m_K^2 +
(c_2+c_5-c_1-2c_6)M_\phi^2+(c_1-c_2+c_5)s \,] \,,
\nonumber \\&&
\nonumber \\
&a_1 =& \frac{1}{2\sqrt2 M_\phi F_K} \bigg\{\,\, F_V( m_K^2-M_\phi^2-s )
-2G_V( m_K^2+M_\phi^2-s )+ \nonumber \\ && + 2\sqrt2 F_A [\, \lambda_0 m_K^4  +
( M_\phi^2-s )( \lambda' M_\phi^2 -\lambda'' s ) -
m_K^2 (\lambda_0 M_\phi^2 +\lambda' M_\phi^2 +\lambda_0 s+\lambda'' s ) \,] \nonumber \\ &&
\times [\,c_\theta^2  D_{K1H}(s)+ s_\theta^2  D_{K1L}(s)   \,]     \,\,\bigg\}\,,
\nonumber \\&&
\nonumber \\
&a_2 = &-\frac{\sqrt2 G_V M_\phi}{ F_K} D_{K}(s) +\frac{ 2 F_A M_\phi}{ F_K} (\lambda'+\lambda'')
[\,c_\theta^2  D_{K1H}(s)+ s_\theta^2  D_{K1L}(s)   \,]\,,
\nonumber \\&&
\nonumber \\
&a_3 =&\frac{F_V- 2G_V}{\sqrt2 F_K M_\phi} - \frac{\sqrt2 G_V M_\phi}{ F_K} D_{K}(s)+\nonumber \\ &&
 +\frac{ 2 F_A }{ F_K M_\phi} (\lambda_0 m_K^2 +\lambda'' M_\phi^2-\lambda'' s )
[\,c_\theta^2  D_{K1H}(s)+ s_\theta^2  D_{K1L}(s)   \,]\,.
\eqa

\item $\tau^-(p) \rightarrow  \pi^-(p_1) \overline{K}^{*0}(p_2) \nu_\tau(q)$:
\bqa
&v =&-i\frac{\sqrt2 F_{V_1} }{F M_{K^*}}
[\,d_m m_\pi^2 + d_M M_{K^*}^2 + d_s s \,] D_{{K^*}'}(s)
\nonumber\\ &&
 -i\frac{ 2\sqrt2 F_V }{F M_{K^*}} [\,(d_1 + 8d_2)m_\pi^2
 + d_3 (-m_\pi^2 +M_{K^*}^2 +s) \,] D_{K^*}(s)\nonumber\\ &&
+\,i\frac{ 2}{F M_V M_{K^*}} [\, (c_1 + c_2+8c_3-c_5)m_\pi^2
+ (c_2+c_5-c_1-2c_6)M_{K^*}^2+(c_1-c_2+c_5)s \,] \,,
\nonumber \\&&
\nonumber \\
&a_1 =& -\frac{1}{2\sqrt2 M_{K^*} F} \bigg\{\,\, F_V(
m_\pi^2-M_{K^*}^2-s ) -2G_V( m_\pi^2+M_{K^*}^2-s )+ \nonumber \\ &&
+ 2\sqrt2 F_A [\, \lambda_0 m_\pi^4  + ( M_{K^*}^2-s )( \lambda'
M_{K^*}^2 -\lambda'' s ) - m_\pi^2 (\lambda_0 M_{K^*}^2 +\lambda'
M_{K^*}^2 +\lambda_0 s+\lambda'' s ) \,] \nonumber \\ && \times
[\,c_\theta^2  D_{K1H}(s)+ s_\theta^2  D_{K1L}(s)   \,]
\,\,\bigg\}\,, \nonumber \\&&
\nonumber \\
&a_2 =& \frac{\sqrt2 G_V M_{K^*}}{ F} D_{K}(s)
-\frac{ 2 F_A M_{K^*}}{ F} (\lambda'+\lambda'')
[\,c_\theta^2  D_{K1H}(s)+ s_\theta^2  D_{K1L}(s)   \,]\,,
\nonumber \\&&
\nonumber \\
&a_3 =&-\frac{F_V- 2G_V}{\sqrt2 F M_{K^*}}
+\frac{\sqrt2 G_V M_{K^*}}{ F} D_{K}(s)-\nonumber \\ &&
 -\frac{ 2 F_A }{ F M_{K^*}} (\lambda_0 m_\pi^2 +\lambda'' M_{K^*}^2-\lambda'' s )
[\,c_\theta^2  D_{K1H}(s)+ s_\theta^2  D_{K1L}(s)   \,]\,.
\eqa
\end{itemize}

The form factors for the $\Delta S =0$ processes are:
\begin{itemize}
\item $\tau^-(p) \rightarrow  \pi^-(p_1) \rho^0(p_2) \nu_\tau(q)$
\bqa
v=0 \,,
\eqa
\bqa &a_1 = &\frac{1}{2 M_\rho F} \bigg\{\,\, F_V(
m_\pi^2-M_\rho^2-s )-2G_V( m_\pi^2+M_\rho^2-s )+2\sqrt2 F_A  D_{a_1}(s)\times
\nonumber \\ &&
\times \big[\, \lambda_0 m_\pi^4  + ( M_\rho^2-s )( \lambda'
M_\rho^2 -\lambda'' s ) - m_\pi^2 (\lambda_0 M_\rho^2 +\lambda'
M_\rho^2 +\lambda_0 s+\lambda'' s ) \,\big]\,\bigg\} \,, \nonumber
\\&&
\nonumber \\
&a_2 =& -\frac{ 2 G_V M_\rho}{ F} D_{\pi}(s)
+\frac{  2\sqrt2 F_A M_\rho}{ F} (\lambda'+\lambda'')D_{a_1}(s) \,,
\nonumber \\&&
\nonumber \\
&a_3 =&\frac{F_V- 2G_V}{ F M_\rho} - \frac{2 G_V M_\rho}{ F} D_{\pi}(s)
 +\frac{  2\sqrt2 F_A }{ F M_\rho} (\lambda_0 m_\pi^2 +\lambda'' M_\rho^2-\lambda'' s ) D_{a_1}(s) \,.
\eqa

\item $\tau^-(p) \rightarrow \pi^-(p_1)\omega(p_2) \nu_\tau(q)$
\bqa
v =&&
-i\frac{2 F_{V_1} }{F M_\omega} [\,d_m  m_\pi^2 + d_M  M_\omega^2 + d_s s \,] D_{\rho'}(s)
\nonumber\\ &&
-i\frac{4 F_V }{F M_\omega} [\,(d_1 + 8d_2)m_\pi^2 + d_3 (-m_\pi^2 +M_\omega^2 +s) \,] D_{\rho}(s)
\nonumber\\ &&
+i\frac{ 2\sqrt2}{F M_V M_\omega} [\, (c_1 + c_2+8c_3-c_5)m_\pi^2
+ (c_2+c_5-c_1-2c_6)M_\omega^2+(c_1-c_2+c_5)s \,]\,,
\nonumber \\
\eqa
\bqa
a_i =0 \,.
\eqa

\item $\tau^-(p) \rightarrow K^-(p_1) K^{*0}(p_2) \nu_\tau(q)$
\bqa
&v =&- i\frac{2\sqrt2 F_V }{F_K M_{K^*}} [\,(d_1 + 8d_2)m_K^2
+ d_3 (-m_K^2 +M_{K^*}^2 +s) \,] D_{\rho}(s) \nonumber\\ &&
+\,i\frac{ 2}{F M_V M_{K^*}} \bigg[\, (c_1 + c_2+8c_3-c_5)m_K^2
+ (c_2+c_5-c_1-2c_6)M_{K^*}^2+\nonumber \\&&
+(c_1-c_2+c_5)s \,\bigg] -i\frac{\sqrt2 F_{V_1} }{F_K M_{K^*}}
[\,d_m  m_K^2 + d_M  M_{K^*}^2 + d_s s \,]
D_{\rho'}(s)\,,
\nonumber \\&&
\nonumber \\
&a_1 =& -\frac{1}{2\sqrt2 M_{K^*} F_K} \bigg\{\,\, F_V( m_K^2-M_{K^*}^2-s )
 -2G_V( m_K^2+M_{K^*}^2-s ) +\nonumber \\ &&
+ 2\sqrt2 F_A [\, \lambda_0 m_K^4  +
( M_{K^*}^2-s )( \lambda' M_{K^*}^2 -\lambda'' s )\nonumber \\&&
 - m_K^2 (\lambda_0 M_{K^*}^2 +\lambda' M_{K^*}^2 +\lambda_0 s+\lambda'' s ) \,]
D_{a_1}(s) \,\bigg\} \,,
\nonumber \\&&
\nonumber \\
&a_2 =& \frac{ \sqrt2 G_V M_{K^*}}{ F_K} D_{\pi}(s) -\frac{  2 F_A M_{K^*}}{ F_K}
(\lambda'+\lambda'') D_{a_1}(s)\,,
\nonumber \\&&
\nonumber \\
&a_3 =&\frac{-F_V+ 2G_V}{\sqrt2 F_K M_{K^*}} + \frac{ \sqrt2 G_V M_{K^*}}{ F_K} D_{\pi}(s)
-\frac{  2 F_A }{ F_K M_{K^*}}
(\lambda_0 m_K^2 +\lambda'' M_{K^*}^2-\lambda'' s ) D_{a_1}(s)\,, \nonumber \\
\eqa
\end{itemize}
where the contributions from the new vector multiplet $V_1$
have been included in the expressions above. We have used the same notation $a_1$
for the axial vector resonance $a_1(1260)$ and the axial vector form factor. This
should not cause any confusion.

\end{document}